\documentclass[12pt,a4paper]{article}
\usepackage{graphicx}
\usepackage{setspace}
\usepackage{natbib}
\usepackage{amssymb}
\usepackage{epsfig,color}
\usepackage{fancyhdr}
\usepackage{amsmath}
\usepackage{float}
\usepackage{subfig}

\renewcommand{\log}{\mathop{\mathrm{lg}}\nolimits}
\newcommand{\cc}{~cm$^{-3}$}   
\newcommand{\sqc}{~cm$^{-2}$}   

\begin{document}

\pagestyle{plain}

\flushright{
{\bf \emph{A talk, presented at \\
the $10^{\rm th}$ Serbian-Bulgarian Astronomical Conference\\
Belgrade, Serbia, May 30 -- June 3, 2016}}
}
\vspace*{24pt}
\begin{center}
{\LARGE \bf Density scaling relation in Orion A:\\ effects of region selection\vspace*{18pt}\\}
{\Large Orlin Stanchev$^{\,1}$, Todor V. Veltchev$^{\,1,\,3}$, \& Sava Donkov$^{\,2}$\vspace*{10pt}}
\flushleft{\large \it
$^1$ Faculty of Physics, University of Sofia, Bulgaria \\
$^2$ Department of Applied Physics, Technical University, Sofia, Bulgaria \\
$^3$ Institute of Theoretical Astrophysics, Heidelberg, Germany \vspace*{6pt}\\}
{\footnotesize {\bf E-mails:}~o\_stanchev@phys.uni-sofia.bg, eirene@phys.uni-sofia.bg, savadd@tu-sofia.bg \vspace{12pt}\\}
\end{center}

\begin{quote}
{\bf Abstract:} Recently Stanchev et al. (2015) proposed a technique to derive density scaling relations in a star-forming region from analysis of the probability distribution function of column density. We address the possible dependence of the outcome on the selection of probe zones, applying the method to {\it Planck} dust-opacity data on Orion A. The derived steep scaling relation of mean density with index $-1.6$ in the molecular cloud (so called `Central filament') points to its self-gravitating nature. The result is reproduced also for large parts of the clouds' vicinity which indicates major role of gravity in the energy balance of the entire star-forming region. 
\end{quote}

\flushleft

\section{Introduction}
Supersonic turbulence has been recognized as a key factor in the star-formation process. Its ubiquity in Galactic star-forming regions (SFRs) accounts for their fractal structure and is testified by scaling relations of velocity dispersion and mean density (or mass), found by numerous authors \citep[][etc.]{Larson_81, Solomon_ea_87, Heyer_ea_09}. The interplay between turbulence and gravity with the evolution of SFRs can be studied by analysis of the probability distribution function of column density $N$ (hereafter, $N$-pdf). The shape of the latter varies from nearly lognormal to a power law while in the most cases it is some combination of both  \citep{Kainulainen_ea_09, LAL_11, Schneider_ea_13}. A possible physical interpretation is that a purely lognormal pdf of {\it spatial} density is indicative for isothermal turbulent medium with negligible self-gravity \citep{VS_94} which transforms -- under conditions of point symmetry -- into a lognormal $N$-pdf. A power-law (PL) `tail' at high densities should develop at timescales of order of the free-fall time, as demonstrated in several numerical works \citep[e.g.][]{Klessen_00, Kritsuk_ea_11, FK_13, Girichidis_ea_14}. \vspace*{6pt}
\\ 
In a purely turbulent medium, a lognormal (column-)density pdf should be derived considering each spatial scale within the inertial range, i.e. well below the scale of energy injection and far above the dissipation scale. Then, if the observational $N$-pdf of a SFR is decomposed into a series of lognormal components, the latter could be interpreted as representative for typical column densities, corresponding to different spatial scales. Led by this idea, \citet{Stanchev_ea_15} derived density scaling relations in the Perseus SFR through analysis of column-density distributions in a number of zones, including the molecular cloud Perseus, as well in its diffuse vicinity. One caveat of the proposed method is the results' possible sensitiveness to the selection of zones: their shape and location, their star-forming activity, balance between gravitational and turbulent energy in them, effects of distance gradient within the chosen SFR. We address some of those problems in the present report. \vspace*{6pt}

\section{Method and research approach}

\subsection{The Orion A star-forming region}
The object chosen for this study is the Orion A complex: probably the most intensively investigated nearby SFR. It is suitable for our purposes due to its location far from the Galactic plane -- its column-density map might be not significantly contaminated by fore- or background cloudy structures (Fig. \ref{fig_Regions_in_OrionA}). Moreover, its physics and structure offer excellent opportunities to probe different effects of zone selection on the $N$-pdf:

\begin{figure}[ht]
	\begin{center}
		\includegraphics[angle=0, width=0.6\textwidth]{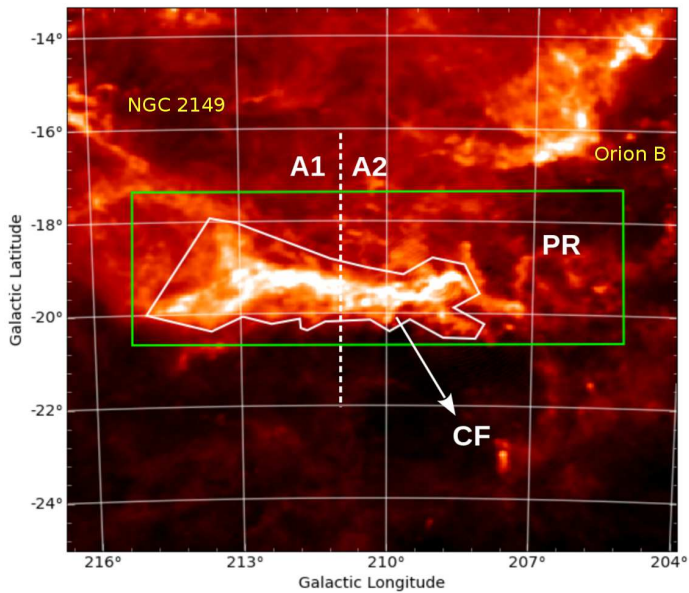}
		\caption{Component map of the dust opacity at frequency 353 GHz, extracted from the {\it Planck} archive. The chosen frame of consideration (green line), the delineated central filament (CF; white), the polygonal ring (PR) around it and the zones of different regimes of star formation (A1, A2) are indicated -- see Sect. \ref{Selection of zones}.}
		\label{fig_Regions_in_OrionA}
	\end{center}
\end{figure}

\begin{figure}[ht]
	\begin{center}
		\includegraphics[angle=0, width=0.95\textwidth]{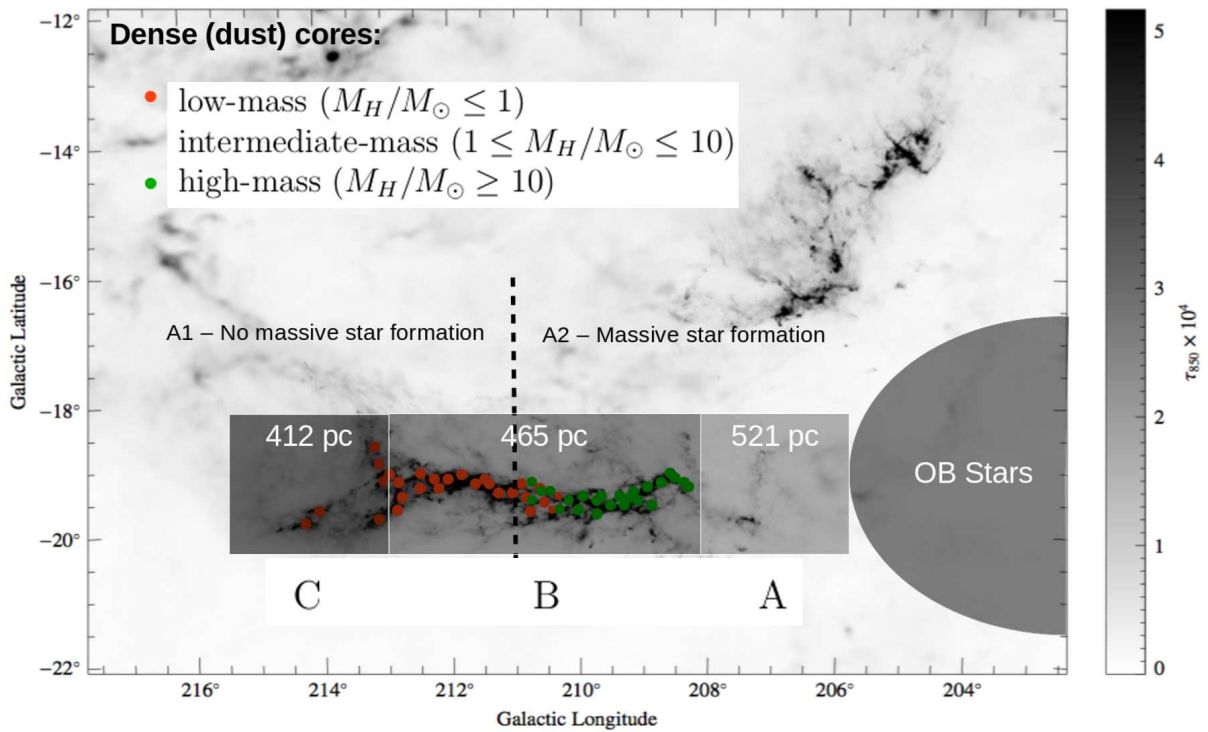}
		\caption{Distributions of dense cores, imposed on {\it Planck}/{\it Herschel} optical-depth map of \citet{Lombardi_ea_14}. The regions A, B, C with different mean distances are shown \citep{Wilson_ea_05}. The spot, labelled `OB stars', denotes the densest part of the Ori OB 1 association. }
		\label{fig_OrionA_characteristics}
	\end{center}
\end{figure}

\begin{itemize}
 \item The giant molecular cloud, often called `Central filament', is elongated and clearly distinguishable from the diffuse environment due to its high mean column density. Most of its substructures are coherent in the velocity space \citep{Bally_ea_87}. The cloud seems to undergo global gravitational collapse \citep{HB_07} and gravitational energy might dominate over the turbulent one at all scales down to subparsec structures \citep{Li_Burkert_16b}.
 \item The star-forming activity in the Central filament could be traced by use of extensive published catalogs of young stellar objects \citep{Megeath_ea_12} and of dense dust/C$^{18}$O cores \citep{Shimajiri_ea_15}. The statistics of massive and bound cores allows for delineation of zones where recent star formation takes place -- see the locations of dust cores in Fig. \ref{fig_OrionA_characteristics}. 
 \item A distance gradient $\Delta D\sim 100$~pc has been discovered along the Central filament, from north to south (\citealt{Wilson_ea_05}; see Fig. \ref{fig_OrionA_characteristics}). It is comparable to the independently estimated distance $D\simeq371$~pc to Orion A SFR \citep{LAL_11} and can therefore significantly affect the scaling relations.
\end{itemize}

\subsection{Observational data and selection of zones}
\label{Selection of zones}
A component map of dust opacity at frequency 353 GHz was extracted from the {\it Planck} archive\footnote{Freely accessible via the Planck Legacy  Archive  interface: http://pla.esac.esa.int/}. The frame of consideration was chosen to contain the Orion A SFR but to exclude Orion B, Mon R2 (twice more distant) and the region NGC 2149 which differs kinematically from the Central filament \citep{Wilson_ea_05}. Following \citet{Stanchev_ea_15}, we prefer to use for our analysis dust-emission instead of dust-extinction data for two reasons: i) to avoid selection effects related to detection of background stars; and ii) to achieve better resolution in zones of higher column density which is crucial for a proper description of the PL-tail regime.\vspace*{6pt}

The transformation from dust optical depth $\tau_{\rm 353}$ to hydrogen column density $N_{\rm H}$ was made by use of the relation of dust opacity to $N_{\rm H}$ as derived in \citet[][see Fig. 20a there]{Planck_XI_13}. The latter was parametrized by a combination of a power-law fit for column densities $N_{\rm H}\le1.5\times10^{21}$~cm$^{-2}$ and a linear fit above this value. Possible non-linearity for optical depths $\tau_{\rm 353}\gtrsim4\times10^{-4}$, reported by \citet{Lombardi_ea_14}, was not accounted for, in view of the poor pixel statistics in this range. \vspace*{6pt}

The selection of zones for our research is illustrated in Fig. \ref{fig_Regions_in_OrionA}. The zone CF encompasses the Central filament, including the integral-shape filament with the Orion Nebula Cluster, the dark cloud L 1641 and other sites of star formation \citep[see][]{Bally_ea_87}. The molecular-emission maps of this region reveal numerous clumps, filaments and shells. \citet{HB_07} proposed a model of evolved self-gravitating disk whose outlook resembles the CF. This inspires us to test the hypothesis that the Orion A SFR -- similar to Perseus SFR (see \citealt{Stanchev_ea_15}) -- can be divided in two spatial domains, governed by different physical regimes: 

\begin{itemize}
 \item {\it Gravoturbulent}, characterized by some equipartition between the specific gravitational and turbulent (kinetic) energy;
 \item {\it Predominantly turbulent}, associated mainly with the diffuse neighbourhood of Orion GMC.
\end{itemize}

In view of this objective, we chose the boundary of CF to follow approximately the sharp-contrast line between zones of high and low column density (Fig. \ref{fig_Regions_in_OrionA}). Additional criterion for selection of zones is the evidence for massive star formation. We adopt the division of \citet{Nishimura_ea_15}: the line $l=211^\circ$ divides the central filament structure in subzones CF1 and CF2 and the polygonal ring, outlined by the boundaries of CF and the whole frame of consideration, in subzones PR1 and PR2. 

\subsection{Decomposition of the $N$-pdf}
A detailed description of the procedure for decomposition of the $N$-pdf is given in the work of \citet{Stanchev_ea_15}, to which we refer the reader. The principle idea is that the column-density distribution in any region of diffuse gas can be represented as a combination of several lognormal functions of type

\begin{equation}
\text{lgn}_{i}(N; a_{i}, N_{i}, \sigma_{i}) = \frac{a_{i}}{\sqrt{2\pi\sigma_{i}^{2}}}\exp{\left(-\frac{[\lg(N/N_{i})]^{2}}{2\sigma_{i}^{2}}\right)}
\end{equation}

where $a_{i}, N_{i}$ and $\sigma_{i}$ are the fitting parameters of the decomposition procedure. A possible interpretation of this decomposition is that each lognormal component represents a spatial domain (scale) with typical column density $N_{i}$ and effective size $L_{i}$: 

\begin{equation}
\label{eq_scale_estimation}
L_{i} = \sqrt{\frac{a_{i}}{\sum_{i}{a_{i}}}}R\,\,. 
\end{equation}

\begin{figure}[h]   
	\centering
	\subfloat{\includegraphics[width=0.5\textwidth, keepaspectratio]{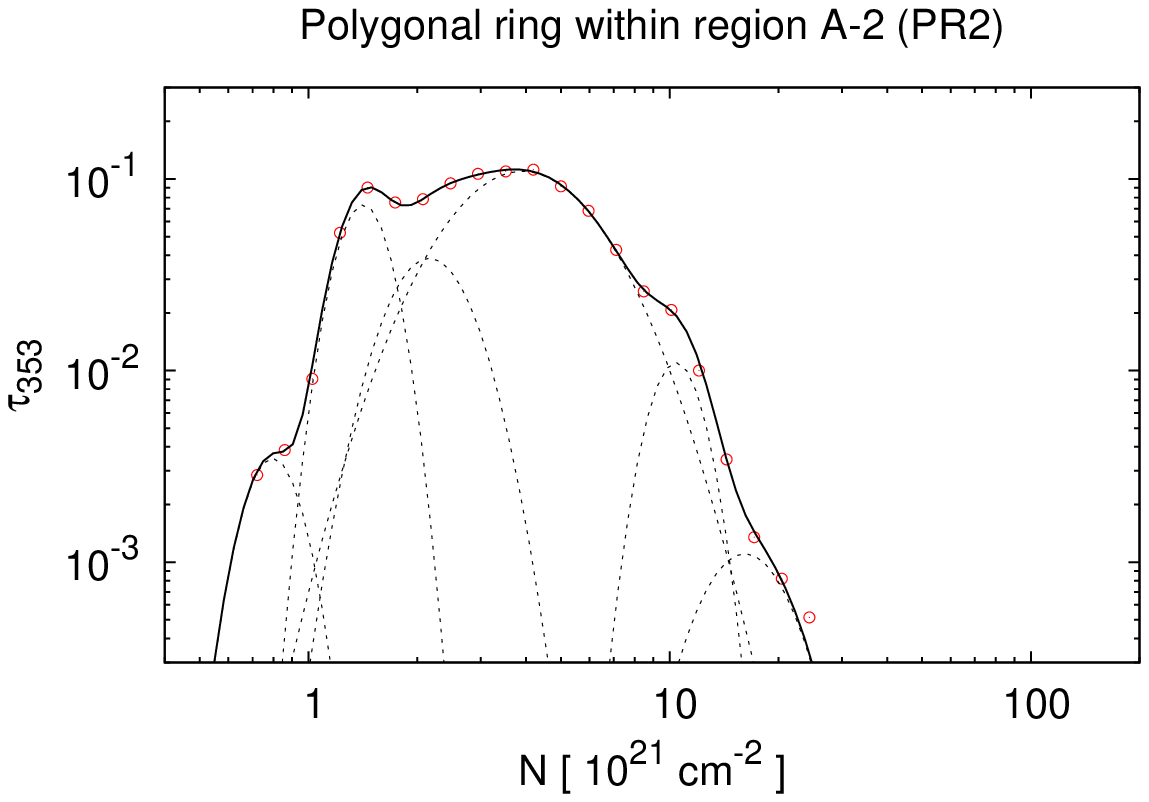}}
	\subfloat{\includegraphics[width=0.5\textwidth, keepaspectratio]{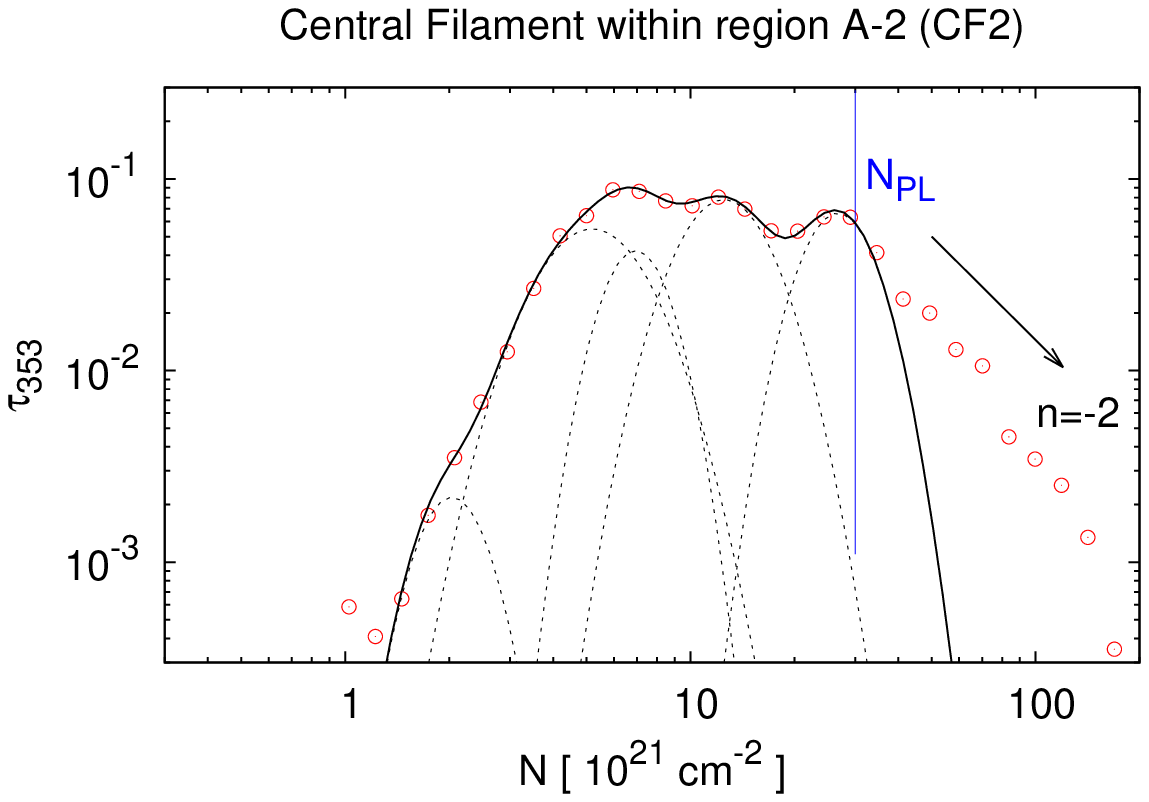}} \vspace{24pt}
	\caption{Column-density pdfs for the zones PR2 (left) and CF2 (right). The dotted curves denote the extracted lognormal components. The deviation point of the PL tail (blue line) and its slope are indicated.}
	\label{fig_pdfs}
\end{figure}

In Fig. \ref{fig_pdfs} are plotted examples of the derived $N$-pdfs: in the zones CF2 and PR2 within the region of active star formation A2. In the former case (left), the $N$-pdf can be described merely by a series of lognormal functions which suggests a predominantly turbulent physical regime. A PL tail emerges additionally in the latter case (Fig. \ref{fig_pdfs}, right), with point of deviation $N_{\rm PL}\sim 3\times10^{21}$\sqc~and slope $n\simeq-2$ as estimated through the method {\sc Plfit} \citep{Clauset_ea_09} from the unbinned observational data. These results are in good agreement with other observations of molecular cloud complexes \citep{FK_13, Kainulainen_ea_09, Schneider_ea_15a} and with numerical studies of self-gravitating clouds \citep{Kritsuk_ea_11}. The derived PL-tail slope $n\simeq-2$ is typical for self-gravitating turbulent media and corresponds to a slope of the volume-density pdf of $-1.5$ \citep[see][]{Girichidis_ea_14, Stanchev_ea_15}. On the other hand, high absolute values of $n$ point to an earlier, predominantly turbulent phase of cloud evolution. Such is the case in the CF1 zone (not shown) where we found $n\simeq-4$. 

\section{Results}

The mean densities of scales $L_i$ which correspond to lognormal $N$-pdf components with typical column densities $N_i$ are obtained straightforwardly:
\begin{equation}
\label{eq_mean_density_lognormal}
\langle n\rangle_{i} = N_{i}/L_{i}~~.
\end{equation}

\begin{figure}[ht]
	\begin{center}
		\includegraphics[angle=0, width=0.9\textwidth]{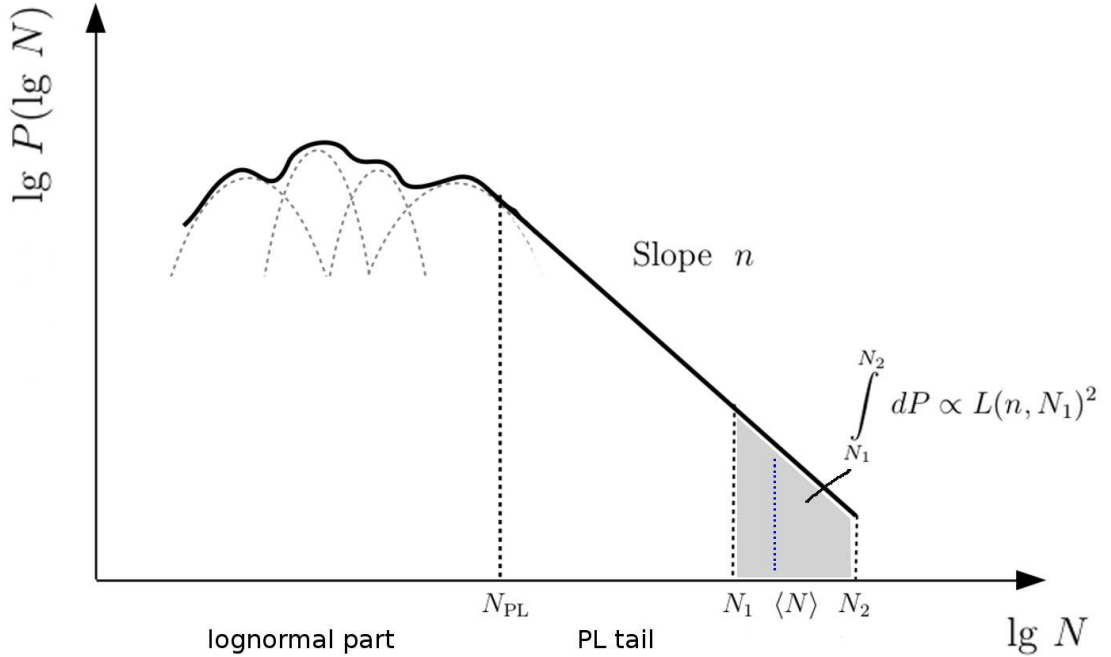}
		\caption{On the estimation of scales and mean column densities in the PL-tail range of the $N$-pdf. See text for notations.}
		\label{fig_pltail_description}
	\end{center}
\end{figure}

To derive the density scaling relation corresponding to the PL tail, one needs first to obtain the deviation point $N_{\rm PL}$ and the total effective size $L_{\rm PL}$ encompassing this column-density range from analysis of the $N$-pdf. Second, the integration between some column density $N_1$ and the upper limit $N_2$ of the PL tail $dP \propto (N/N_{\rm PL})^n$ yields a total area (Fig.~\ref{fig_pltail_description}) which defines an effective size (scale) as a function of $N_1$ and the slope $n$:  

\begin{equation}\label{eq:5}
L(N_1; n) = L_{\rm PL}\left(\frac{N_1}{N_{\rm PL}}\right)^{n/2}~~.
\end{equation}

Finally, the mean density is assessed by dividing the mean column density in the range $[N_1,N_2]$ to its corresponding statistical scale:

\begin{equation}
\label{eq_mean_density_PLT}
\langle n \rangle_L = \frac{\langle N \rangle(N_1, N_2; n)}{L(N_1; n)}~~.
\end{equation} 

We adopt $\langle N \rangle = \langle N \rangle_{\rm log}$ where the `logarithmic average' $\langle N/N_{\rm PL} \rangle_{\rm log}\equiv10^{\overline{\log(N/N_{\rm PL})}}$ is calculated by use of the formalism described in \citet{DSV_12}. In the case of a long PL tail ($N_2/N_1\gg1$), it can be demonstrated that $\langle N \rangle_{\rm log}\simeq\exp(-1/n)(N_1)$. The obtained $N$-pdfs in Orion A have short tails (cf. Fig. \ref{fig_pdfs}, right) which introduces a weak dependence of $\langle N \rangle$ on the $N_2$ too.
\vspace*{6pt}

\subsection{Assesing the effect of different physical regimes}

\begin{figure}[ht]   
	\centering
	\subfloat{\includegraphics[width=0.5\textwidth, keepaspectratio]{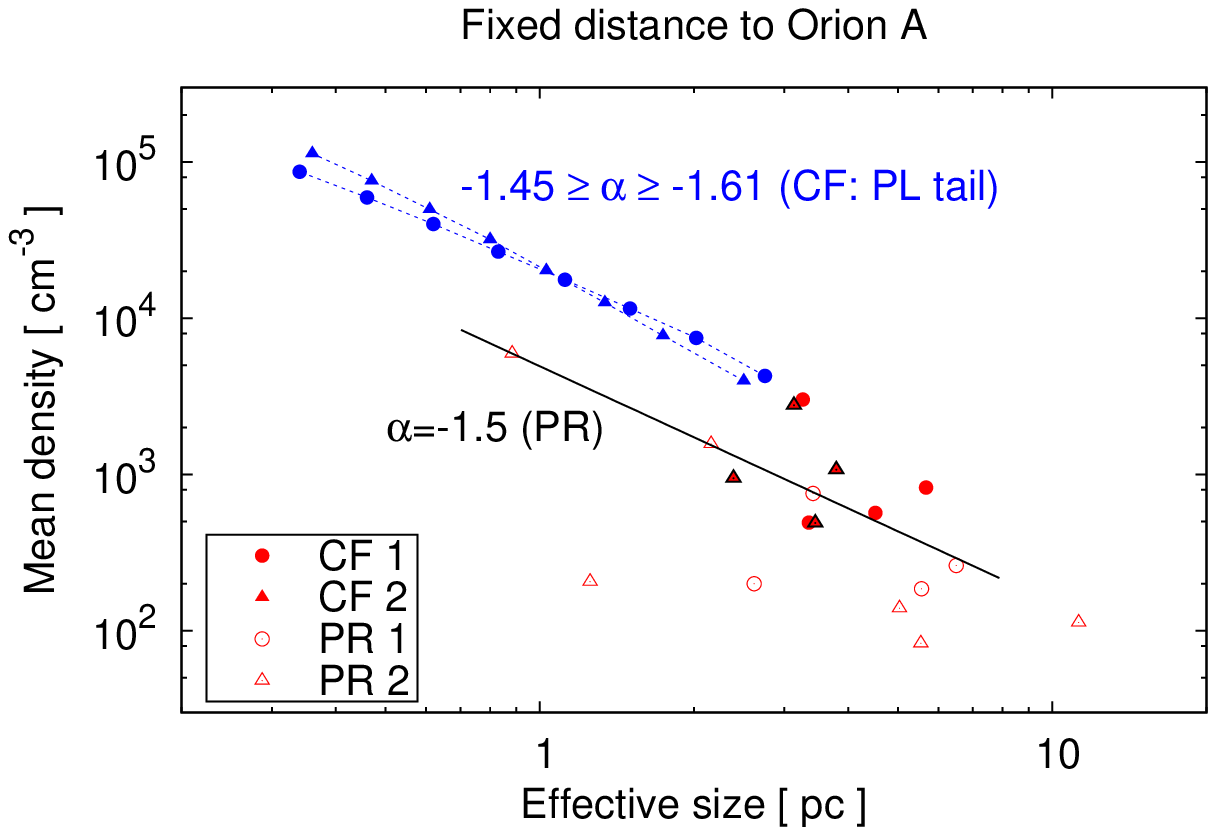}}
	\subfloat{\includegraphics[width=0.5\textwidth, keepaspectratio]{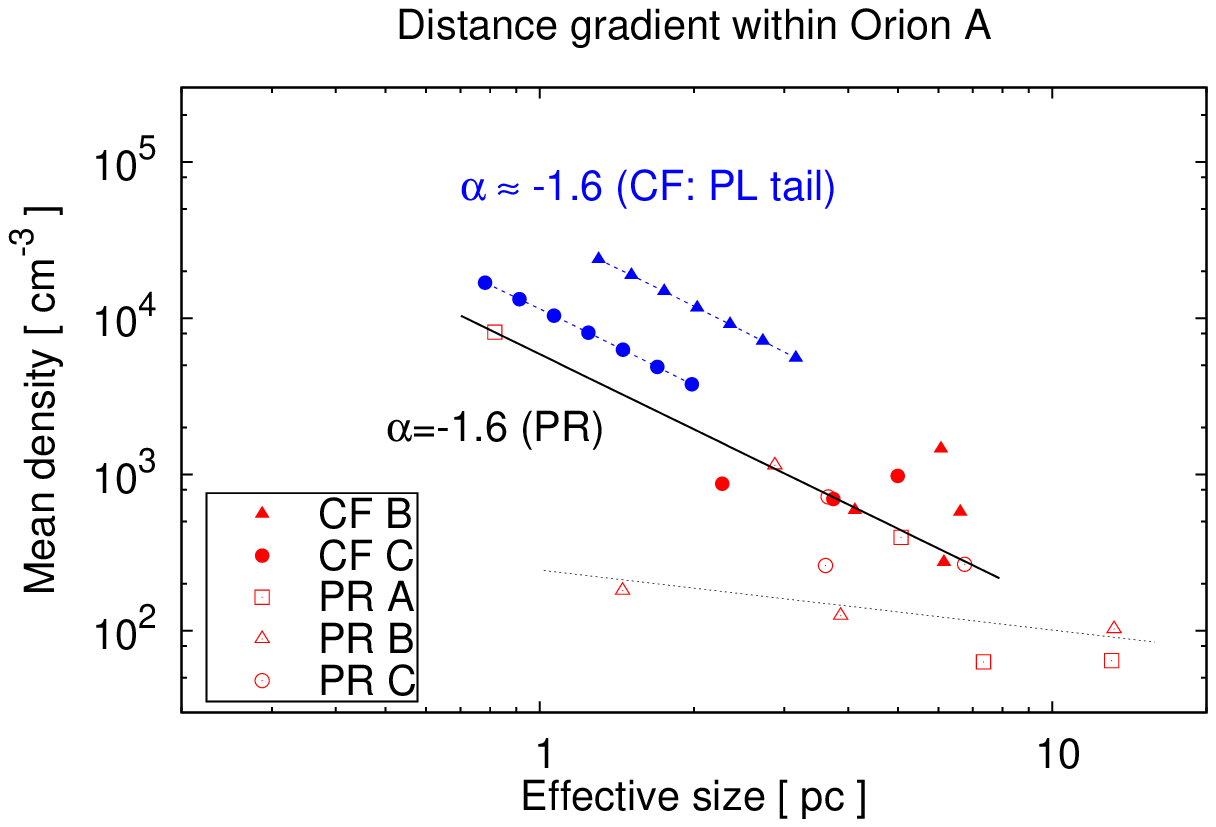}} \vspace{24pt}
	\caption{Density scaling relations $\langle n\rangle \propto L^\alpha$ for a fixed distance to Orion A (left) and taking into account a distance gradient (right). The symbols for the scales in the PL tail regime (blue) denote the same zones like in the lognormal case. Possible scaling for the diffuse components is shown in the right panel (dashed line). See text.}
	\label{fig_density_scaling_relations}
\end{figure}

The density scalings derived from analysis of the lognormal part of the $N$-pdfs (Eq. \ref{eq_mean_density_lognormal}) and of their PL tails (Eq. \ref{eq_mean_density_PLT}) are plotted in Fig. \ref{fig_density_scaling_relations} with red and blue symbols, respectively. The results in the left panel are obtained for a fixed distance $D=371$~pc to Orion A \citep{LAL_11}. As mentioned in Sect. \ref{Selection of zones}, the selection of probe zones was strongly motivated by the model of rotating, self-gravitating disk of \citet{HB_07}. According to the model predictions, the disk undergoes a large-scale gravitational collapse at the final stage of its dynamical evolution when its morphology resembles very much the Central filament in Orion A (see Fig. 6 in \citealt{HB_07}). Therefore we tentatively consider the CF zone as a gravoturbulent domain, characterized by some equipartition between the specific gravitational and turbulent energy, and the PR zone (cf. Fig. \ref{fig_Regions_in_OrionA}) as a predominantly turbulent domain to test whether those different physical regimes affect the density scaling relations.  
\vspace*{6pt}

Considering only the lognormal components, there seems to be two patterns (Fig. \ref{fig_density_scaling_relations}, left). On the one hand, no density scaling is evident for structures with $\langle n\rangle\lesssim200$~\cc. On the other hand, the denser structures in the zones CF and PR exhibit a pronounced size-density relationship, although with significant scatter. The obtained scaling with index $\alpha\sim-1.5$ is steeper than the `classical' second Larson's relation \citep[$\alpha\sim-1$;][]{Larson_81} and much steeper than the one found in the diffuse vicinity of Perseus SFR ($\alpha\simeq-0.8$) by use of the same method \citep{Stanchev_ea_15}. Such behaviour is hard to attribute to structures shaped merely by turbulence. Bringing into consideration CF structures from the PL tail regime (blue) sheds light on the issue -- they obey the same scaling law! Their high mean densities and their association with sites of recent star formation testified by the presence of dust cores and young stellar objects \citep{Megeath_ea_12, Shimajiri_ea_15} suggest that gravity plays a crucial role for their physical characteristics. One is drawn to the conclusion that the density scaling in the Central filament of Orion A and, partially, in its diffuse vicinity has a gravoturbulent origin.     

\subsection{Assessing the effect of distance gradient in Orion A}  
It is necessary to check whether the presented results are affected by the distance gradient, found by \citet{Wilson_ea_05}. Following their work, we divide the Orion A SFR into three subregions with different distances (Fig. \ref{fig_OrionA_characteristics}). The CF zone is covered only by the subregions B and C. As evident from Fig. \ref{fig_density_scaling_relations}, right, the main result does practically not change: a steep density scaling with index $\alpha\approx-1.6$ is derived both from denser lognormal components and from the PL tail, for the CF and the PR zone as well. A weak scaling with index $\alpha_{\rm dif}\simeq-0.4\pm0.2$ is obtained also for the more diffuse components of the PR. 

\section{Discussion and conclusions}

In this work we assess some effects of region selection on the mean density scaling relation, derived by the $N$-pdf decomposition technique. The latter was proposed by \citet{Stanchev_ea_15} and applied first to the Perseus star-forming region. Two different physical domains have been distinguished in Perseus: i) gravoturbulent, encompassing regions which include the molecular cloud and characterized by a power-law $N$-pdf; and, ii) predominantly turbulent, encompassing ring zones around the molecular cloud (excluding the cloud itself) and characterized by a $N$-pdf, which could be decomposed to a series of lognormals. These regimes have different imprint on the density scaling -- the latter is very steep ($\alpha\simeq-2$) in the gravoturbulent case and shallow ($\alpha\simeq-0.8$) in the other one. Making use of the numerous investigations on Orion A, we aimed to elaborate the approach, applying it to {\it Planck} data on this region. A presumably gravoturbulent zone CF was selected which includes the Central filament, resembles the shape of the self-gravitating disk model of \citet{HB_07} and has an effective size $11$~pc, about the largest scale at which gravitational energy may dominate over the turbulent one \citep{Li_Burkert_16b}. A large polygonal zone (PR) around the CF was expected to bear signature of predominant turbulence. \vspace*{6pt}

In contrast to Perseus, the $N$-pdf of the CF in Orion A can be decomposed to a power-law part, combined with a series of lognormals spaning more than an order of magnitude of $N$. The latter components correspond to less dense structures which probably have a turbulent origin.  However, they turn out to obey the same density scaling relation with index $\alpha\simeq-1.6$ like their peers of high column density, represented by the power-law part of the $N$-pdf. The somewhat surpising result is reproduced for denser structures ($\langle n\rangle\gtrsim10^3$~\cc) also in the large diffuse vicinity (PR) of the molecular cloud. The steep density scaling is indicative that gravity plays a major role in the energy balance in the entire star-forming region, at large as well at small scales. It is an independent confirmation of the findings of \citet{Li_Burkert_16a, Li_Burkert_16b} from the gravitational energy spectrum in Orion A, derived in two alternative ways.\vspace*{6pt}
 
The apparent lack of density scaling (or a shallow scaling, Fig. \ref{fig_density_scaling_relations}, right) for less dense structures ($\langle n\rangle\lesssim3\times10^2$~\cc) in the diffuse vicinity lends additional support to the gravoturbulent hypothesis. Such densities are typical for diffuse molecular clouds \citep{Goldsmith_13}. We suggest that the scales in question are short-living, far from a virial-like state.\vspace*{6pt}

To generalize the results of the performed analysis, we conclude that:
\begin{enumerate}
\item  The molecular cloud Orion A, or at least the northern (integral-shaped) part of the Central filament should be self-gravitating since the slope of the power-law tail ($n\simeq-2$) suggests a developed density profile in the high-density zones, typical for self-gravitating cores, and domination of gravity. The steep scaling index of density $\alpha\simeq-1.6$ is also indicative for a gravoturbulent regime rather than for a predominantly turbulent one ($\alpha\simeq-1$). 
\item Large part of the whole star-forming region Orion A is a gravoturbulent domain wherein the gravitational and kinetic energy are in equipartition at multiple scales. This conclusion is supported by the density scaling relation to which obey the denser structures in the large diffuse vicinity of the molecular cloud: identical to the one derived for their peers in the CF ($\alpha\simeq-1.6$).
\item The distance gradient effect does not affect the derived density scaling laws.
\end{enumerate}
 
{\it Acknowledgement:} T.V. acknowledges support by the {\em Deutsche Forschungsgemeinschaft} (DFG) under grant KL 1358/20-1.  


\end{document}